\documentclass[11pt,preprint]{aastex}
\usepackage{epsfig}
\usepackage{graphicx}
\usepackage{url}
\usepackage{natbib}

\shorttitle{Kinematic Discovery of a Stellar Stream located in Pisces}
\shortauthors{Martin et al.}

\begin{document}

\title{Kinematic Discovery of a Stellar Stream located in Pisces}

\author{
Charles Martin\altaffilmark{1},
Jeffrey L. Carlin\altaffilmark{1},
Heidi Jo Newberg\altaffilmark{1},
Carl Grillmair\altaffilmark{2}
}

\altaffiltext{1}{Department of Physics, Applied Physics and Astronomy,
Rensselaer Polytechnic Institute, Troy, NY 12180, USA, martic6@rpi.edu}
\altaffiltext{2}{Spitzer Science Center, California Institute of Technology, 
Pasadena, CA 91125, USA}

\begin{abstract}
We report the kinematic discovery of the Pisces Stellar Stream (PSS), at Galactic longitude 
$l \approx 135^\circ$ and $-39^\circ<b<-36^\circ$. We originally identified this halo substructure 
from velocities of red giant branch stars in Sloan Digital Sky Survey (SDSS) Data Release 8, 
and confirmed its presence in turnoff stars from SDSS photometric data. The PSS is a 
narrow, kinematically cold tidal stream, with $\sigma_{\rm v,0} \approx 8$ km s$^{-1}$. Its metallicity is [Fe/H] $\approx -2.2$, 
with $\sim0.3$-dex dispersion. 
The color-magnitude signature of the stream turnoff, combined with our measured metallicity, places the PSS at a distance of 35$\pm$3 
kpc. The Pisces Stellar Stream is the same as the previously announced 
``Triangulum stream'' and part of the proposed ``stream a''.  We rule out an association of the 
PSS with other previously known Milky Way substructures in the same region of the sky. 

\end{abstract}

	
\section{Introduction}

Many stellar streams and substructures resulting from the tidal disruption of satellite dwarf 
galaxies and star clusters under the gravitational influence of the Milky Way have been identified 
and mapped using deep, large-area photometric surveys 
\citep{2002ApJ...569..245N, 2003MNRAS.340L..21I, 2003ApJ...599.1082M, 2003ApJ...588..824Y, 2004ApJ...615..732R, 2006ApJ...642L.137B,
2006ApJ...639L..17G, 2006ApJ...641L..37G, 2006ApJ...643L..17G,
2006ApJ...651L..29G, 2007ApJ...654..897B, 2009ApJ...693.1118G, 2009ApJ...700L..61N, 2009ApJ...697..207W, 2010gama.conf..247G, 
2010ApJ...712..260K, 2010IAUS..265..255R, 2010ApJ...722..750S, 2012ApJ...757..151L, 2012ApJ...755..134S}. 
Streams and cloud-like substructures have 
also been observed kinematically via spectroscopic surveys of tracers such as RR Lyrae, M-giants, 
or turnoff stars 
(e.g., Vivas et al. 2001; Duffau et al. 2006; Carlin et al. 2012; 
Majewski et al. 2012; Sheffield et al. 2012).

\begin{figure}[!t]
\includegraphics[height=4.0in]{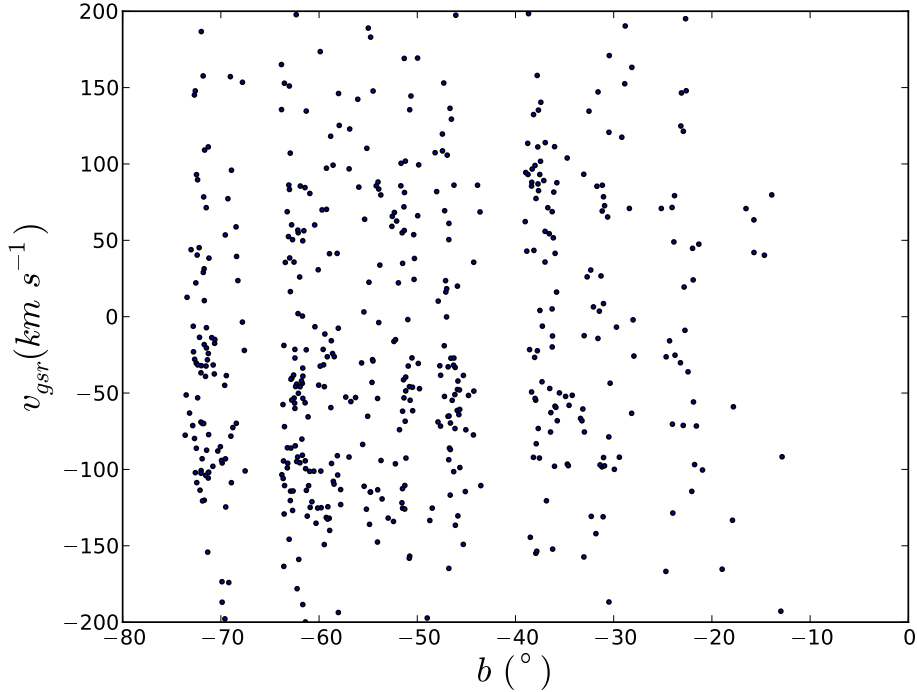}
\caption{Line of sight velocity in the Galactic standard of rest frame, $v_{gsr}$ (calculated using Equation 1 from \citealt{2012ApJ...757..151L}), as a function of Galactic latitude, for stars from SDSS DR8. These are selected as metal-poor 
giant stars 
in the south Galactic cap, 
with $120^\circ<l<160^\circ$. 
Two separate bands of stars with $v_{\rm gsr} < 0$ km s$^{-1}$ sweep across the lower portion of this figure, moving to lower $v_{\rm gsr}$ with increasing latitude. These two structures are associated with the Cetus Polar Stream and the Sagittarius tidal stream. An additional clump of stars is seen at $b\approx -37^\circ$ and $v_{\rm gsr} \approx 90$~km~s$^{-1}$; we show in this work that these stars are part of a stellar stream at a distance of $\sim35$~kpc.}
\label{vgsr_vs_b}
\end{figure}

Here we report a stellar tidal stream discovered kinematically, which (based on its location in the 
constellation Pisces) we dub the ``Pisces Stellar Stream'' (PSS). This stream was identified as a 
grouping of red giant branch stars at a common distance and radial velocity during the course of a 
spectroscopic study of the Cetus Polar Stream (CPS). In the CPS study (Yam et al. 2013), we 
selected metal-poor ($-3.0<$[Fe/H]$<-1.5$) stars with surface gravities consistent with being 
giants ($0<\log~{g}<4.0$) from the SDSS DR8 spectroscopic data in the south Galactic 
hemisphere. The Cetus stream was clearly present in velocity versus Galactic latitude plots of these 
stars, along with an additional feature consisting of a few stars clumped at $v_{\rm gsr}\sim90$~km~s
$^{-1}$. 
Upon further exploration, these stars turn out to nearly all be drawn from the same small region of sky and have 
similar magnitudes (and thus similar distances). In this Letter, we show evidence that this grouping 
of stars is part of a stellar tidal stream at a distance of roughly 35 kpc from the Sun in the 
constellation Pisces. This stream is a piece of ``stream a'' in the matched-filtered density 
maps of Grillmair (2012), and is the same as the ``Triangulum stream'' in Bonaca et al. (2012).
We chose to rename the stream because our detection is within the Pisces constellation, the stream does not pass through Triangulum, and we cannot verify whether the stream extends the length of ``stream a."
This Letter presents the first measurement of velocities and metallicities of stars in this stream. 

\section{Observation and Data Analysis}

\begin{figure}[!t]
\includegraphics[width=5.5in]{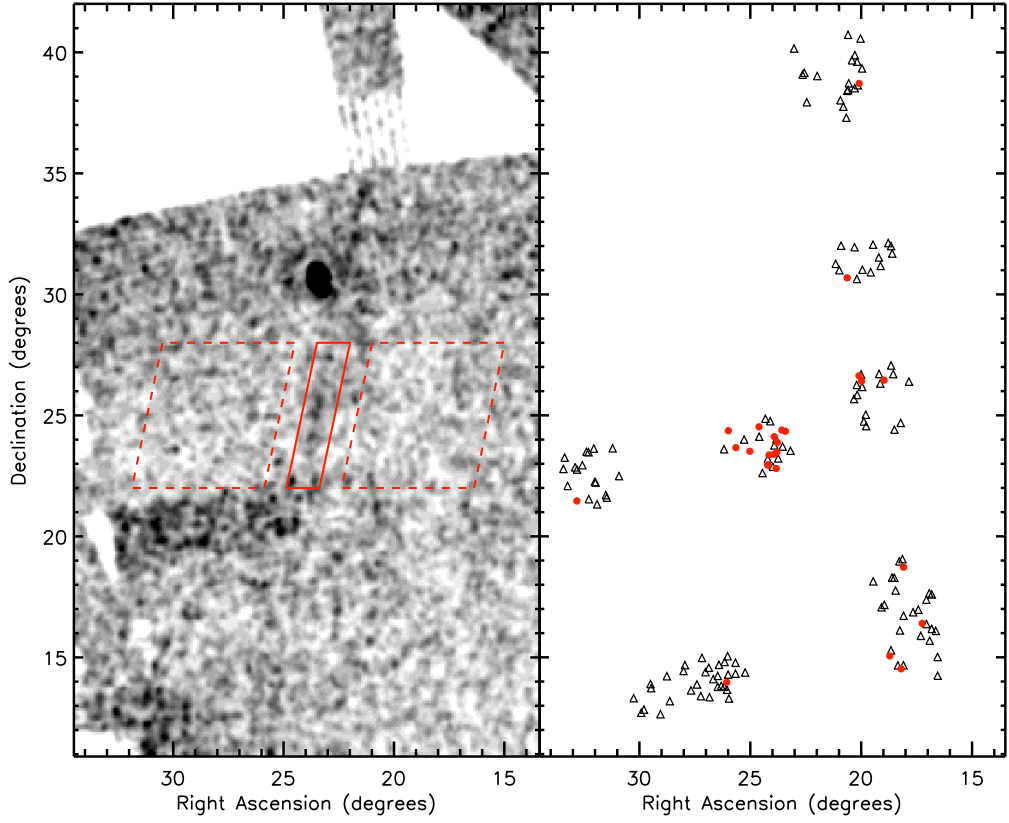}
\caption{{\it Left panel:} map of filtered star counts in the region surrounding the Pisces Stellar Stream; darker shades represent higher densities. This map was created by extracting stars from SDSS DR8 within an isochrone filter of age 12.7~Gyr, metallicity $Z=0.0001$ ([Fe/H]$ \approx  -2.28$), and placed at a distance of 33.6 kpc, 
using the matched filter technique of Grillmair~(2009) with $0.1\arcdeg$ bins. The image was smoothed with a Gaussian kernel three pixels wide. The large ``blob'' at $(\alpha, \delta) \sim (23.5\arcdeg, 30.5\arcdeg)$ is M33. Passing near M33 on the sky is the PSS, which is visible as a narrow feature running diagonally from $(\alpha, \delta) \sim (24\arcdeg, 23\arcdeg)$ to $(21\arcdeg, 35\arcdeg)$, and perhaps further. Red lines and dashed lines show the `on-stream' and `off-stream' regions of the sky, respectively.
Note that the broad feature at $\alpha > 24\arcdeg$ and $19 < \delta < 22\arcdeg$ is a data artifact that arises because of poor seeing when this particular SDSS stripe was observed. {\it Right panel:} Positions of SDSS spectra of metal-poor ($-2.8 < $[Fe/H] $< -1.8$) giant stars (small open triangles). Filled red dots indicate stars with velocities between $75$~km~s$^{-1}$ $< v_{\rm gsr} < 115$~km~s$^{-1}$. The SDSS plate 
centered at $(\alpha, \delta) \sim (24.7\arcdeg, 23.7\arcdeg)$ 
contains most of the velocity-selected red points. Within the plate, these metal-poor, velocity-selected giant stars are mostly clustered in a narrow line of 11 stars. }
\label{starcounts}
\end{figure}

In Figure~\ref{vgsr_vs_b} we 
plot $v_{\rm gsr}$ vs. Galactic latitude for stars between Galactic longitudes of $120^\circ-160^\circ$, $0 < \log
{g} < 4$ (giant stars), $-2.8 <$~[Fe/H]~$< -1.8$, 
$0.4 < (g-r)_0 < 0.9$ and $g_0 > 15.5$.
The color and magnitudes were chosen to exclude both turnoff stars whose surface gravity was miscalculated and evolved post-RGB stars, while retaining likely red giant branch stars.
The most prominent features in Figure~\ref{vgsr_vs_b} are swaths of RGB stars at $v_{\rm gsr} < 0$  
km s$^{-1}$, running across most of this plot, which are associated with the Sagittarius and Cetus streams (Newberg et al.~2009, Koposov et al.~2012, Yam et al.~2013).
An additional 
overdensity of metal-poor giant stars can be seen around $b=-37^\circ$ and $v_{\rm gsr} = 
90$~km~s$^{-1}$. 

We now show that
this velocity clump represents a real structure in the Galactic halo. 
From the sample of metal-poor red giant branch (RGB) stars shown in Figure~\ref{vgsr_vs_b}, we selected a 
smaller velocity range of $75$~km~s$^{-1} < v_{\rm gsr} < 115$~km~s$^{-1}$.
We plotted the positions of stars in Galactic coordinates, and noted that many of these 
velocity-selected giant stars with metallicities between $-2.8 < $[Fe/H]$ < -1.8$ are concentrated in 
a small fraction of a single SDSS spectroscopic plate centered at $l=136^\circ$ and $b=-35^\circ$ (see Figure~\ref{starcounts}).

Figure~\ref{starcounts} was produced using the matched-filter technique described in detail by 
Grillmair~(2009), which is in turn a refinement of the Rockosi~et~al.~(2002) method. 
This method can be used to create weighted density maps of stars from within a given color-magnitude filter 
(based on either an empirical ridgeline or a theoretical isochrone) placed at a given distance; the 
weights emphasize regions of the filter expected to contain the highest ratio of stream stars to foreground stars. 
The map in Figure~\ref{starcounts} shows filtered star counts from SDSS DR8 using a filter based 
on a Padova isochrone (Marigo et al. 2008, Girardi et al. 2010) with $Z=0.0001$ ([Fe/H]$ \sim 
-2.28$), age 12.7 Gyr, at a distance of 33.6 kpc (note that the filter's response will be highest near the
main sequence turnoff of the filter, at colors where foreground contamination is minimal).
The most prominent feature 
(the blob at $\alpha, \delta \sim 23.5^\circ, 30.5^\circ$) is due to the M33 galaxy. A narrow stream is 
visible running nearly vertically across this plot, from $(\alpha, \delta) \sim (24^\circ, 23^\circ)$ to $
(\alpha, \delta) \sim (21^\circ, 35^\circ)$. 

We selected a region of sky based on the positions of these velocity-selected, metal-poor giants that we call the ``on-stream field.'' 
This is defined as the area within $\pm0.75^\circ$ in right ascension about the line $
\delta = -4.33 \alpha + 126.46^\circ$, confined to $22^\circ < \delta < 28^\circ$; the linear trend was 
fit to apparent density peaks in the matched-filter star count map (Figure~\ref{starcounts}). 
The width of the on-stream field was chosen to include all of the spectroscopic candidates we originally identified in this region, and its length 
covers the portion of the sky where the stream is obvious in the filtered star counts map.

An off-stream region was defined that contains an area eight times larger than that of the on-stream 
field. This comparison region was made up of two areas (seen as dashed boxes on Figure~\ref
{starcounts}) defined by:
\begin{equation}
-0.231 \delta + 21.455^\circ  < \alpha < -0.231 \delta + 27.455^\circ, 22^\circ < \delta < 28^\circ;
\end{equation}

\begin{equation}
-0.231 \delta + 30.956^\circ  < \alpha < -0.231 \delta + 36.956^\circ, 22^\circ < \delta < 28^\circ.
\end{equation}

\begin{figure}[!t]
\includegraphics[height=5.75in]{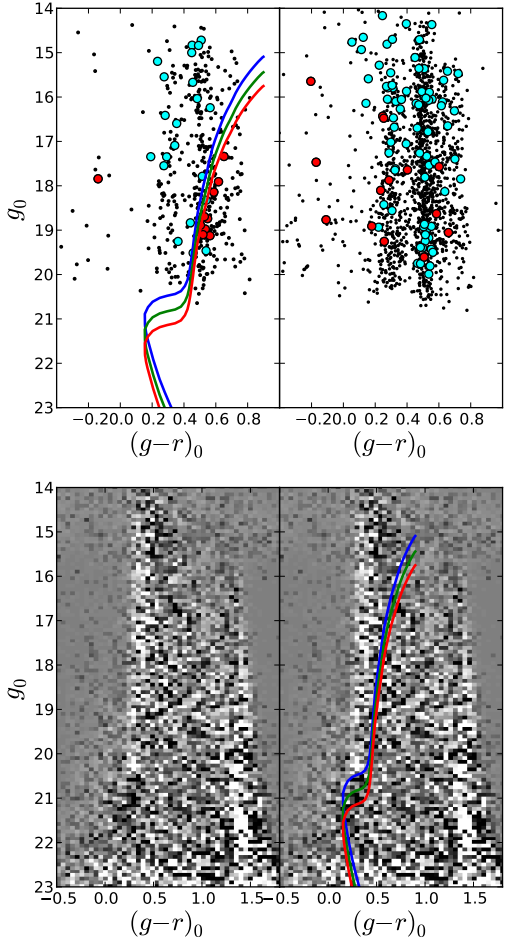}
\caption{\scriptsize {\it Top row:} color magnitude diagrams of stars within the on-stream (upper left panel) and off-stream (upper right panel) fields defined in Section~2.
Small black dots represent all stars with SDSS DR8 spectra. Light blue points in both panels represent giant stars with Galactocentric line-of-sight velocities $75$~km s$^{-1} < v_{\rm gsr} < 115$~km s$^{-1}$. Red points 
have an additional metallicity constraint of
$-2.8 <$~[Fe/H]~$< -1.8$. 
The left (on-stream) panel has a clear concentration of stars grouped at
$0.5 \lesssim (g-r)_0 \lesssim 0.7$, $g_0 \sim 18.8$, 
indicative of a red giant branch. Overplotted Padova isochrones 
correspond to a metallicity of [Fe/H]~$\approx -2.3$ and an age of 12.7 Gyr. Blue, green, and red isochrones correspond to distances of 28, 33, and 38 kpc, respectively. 
{\it Bottom row:} Color magnitude Hess diagrams showing stellar density in the on-stream field after subtraction of the local background (off-stream, scaled down by a factor of eight) contribution. The left panel shows the residuals; the right panel is the same Hess diagram with Padova isochrones from the upper left panel overplotted. 
Darker pixels signify excess on-stream stars. A main sequence turnoff is apparent at $g_0~\sim~21$ in the lower panels. }
\label{spectra_CMD}
\end{figure}

In the top left and top right panels of Figure~\ref{spectra_CMD} we show color magnitude diagrams 
(CMDs) of all stars having SDSS DR8 spectra within the on- and off-stream fields, respectively. We also show subsets of the spectra with properties expected for the PSS.  The 
subscripts on the magnitudes $g_0$ and $(g-r)_0$ mean that these magnitudes have been 
extinction corrected based on the Schlegel et al. (1998) maps (using the extinction values provided 
for each star in the SDSS database). 
 
The metal-poor giant stars selected using the velocity criteria (i.e., red points in the top panels of 
Figure~\ref{spectra_CMD}) are mostly clustered around $[(g-r)_0,g_0] = [0.55, 18.8]$ in the on-stream 
field (top left panel of Figure~\ref{spectra_CMD}), while the more metal-rich stars (light blue points) 
scatter throughout the diagram. We suggest that these stars are part of the red giant branch
of a single structure. The comparison off-stream field (top right panel) contains several stars that 
match our velocity, metallicity, and giant-selection criteria, but don't have a coherent distribution 
that could reasonably correspond to a giant branch. 
In the on-stream field there are a total of 448 
stars with spectra in this diagram: 32 light blue points (velocity-selected giants) and 11 red points 
(velocity {\it and} metallicity-selected giants). The off-stream area has 1584 total 
spectra, with 90 light blue points and 14 red points. To calculate potential foreground contamination we selected stars with $g_0 < 19.2$ that are within 0.1 magnitudes in $g_0, (g-r)_0$ of the Padova isochrone with age 12.7 Gyr, [Fe/H] = -2.3, at a distance of 33 kpc. There are 403 and 124 spectra within this filter in the off- and on-stream regions, respectively; a factor of 3.25 times as many 
spectra were observed in the off-stream region. Among the off-stream spectra in this isochrone filter, there are only 2 metal-poor ($-2.8 < {\rm [Fe/H]} < -1.8$) red giants ($\log g < 4.0$) with $75 < v_{gsr} < 115~{\rm km~s^{-1}}$. Scaling by the factor of 3.25, this leads to an expectation of $\sim1$ background ``interloper'' in our on-stream sample.  Our finding of 10 candidate stream stars in this region (satisfying the same criteria) is thus a significant excess relative to neighboring 
regions. Furthermore, these stars are clustered in a small region of the CMD where we expect RGB stars.
One additional candidate is a blue star that is consistent with being a 
horizontal branch star at the same distance. Table 1 shows the properties of the eleven stars 
classified as PSS candidates.

On the top left panel of Figure~\ref{spectra_CMD} we overplot three isochrones shifted to 
distances (28, 33, and 38 kpc) that roughly match the RGB,
with a metallicity of $Z=0.0001$ ([Fe/H]$\sim -2.28$) and age 12.7 Gyr.
Our choices of isochrone parameters will become clear later in this paper, as these 
were motivated by the main sequence turnoff (MSTO) visible in our background-subtracted 
Hess diagram, our spectroscopic metallicity measurement, and examination of our matched-filter star count map. 
For now, we note that the RGBs of the fainter two isochrones match the clump of red points in Figure~\ref
{spectra_CMD} well.

\begin{figure}[!t]
\includegraphics[width=3.3in, clip=true, trim=1.5cm 0.5cm 0.5cm 1.0cm]{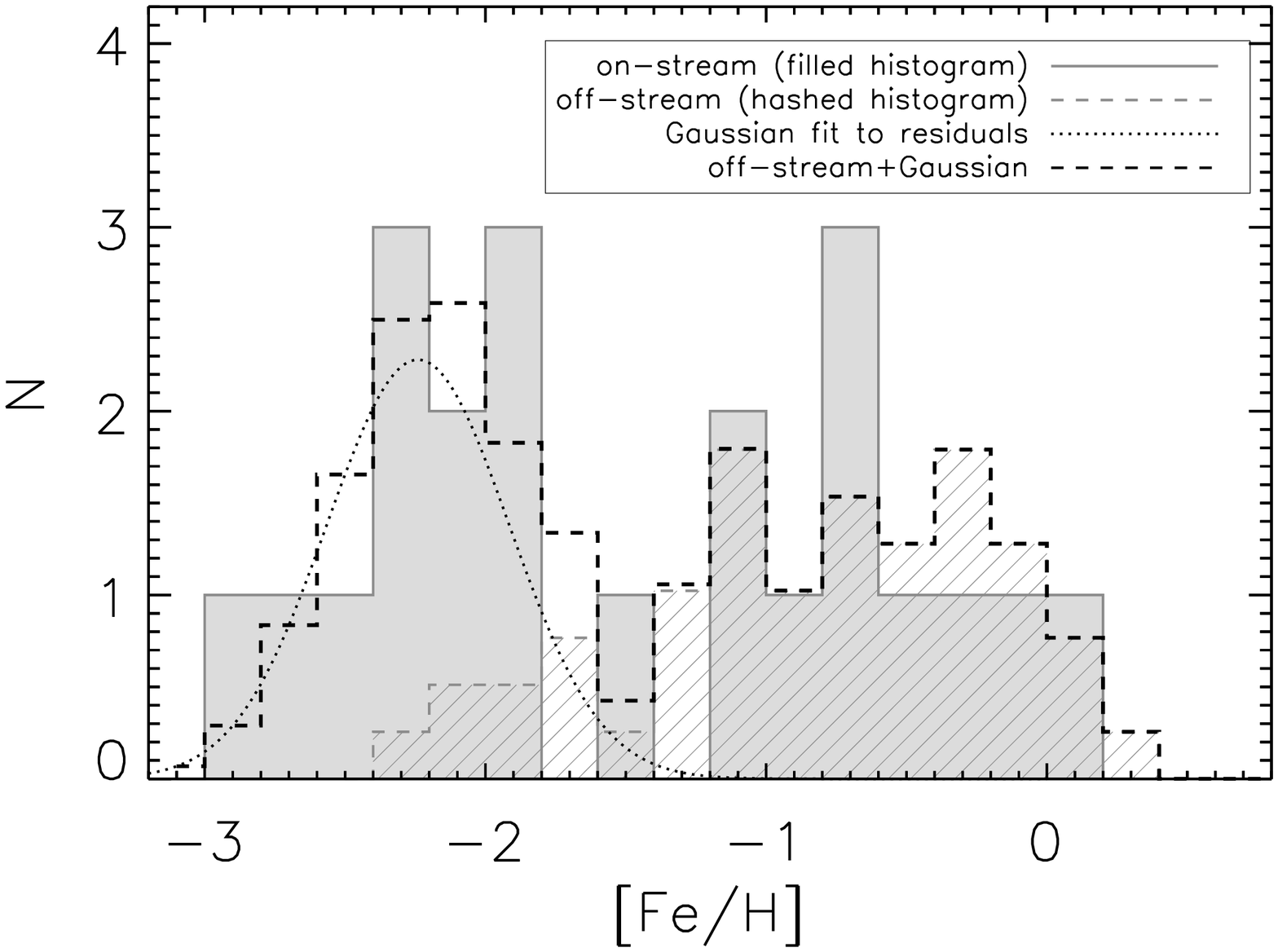}
\includegraphics[width=3.3in, clip=true, trim=1.5cm 0.5cm 0.5cm 1.0cm]{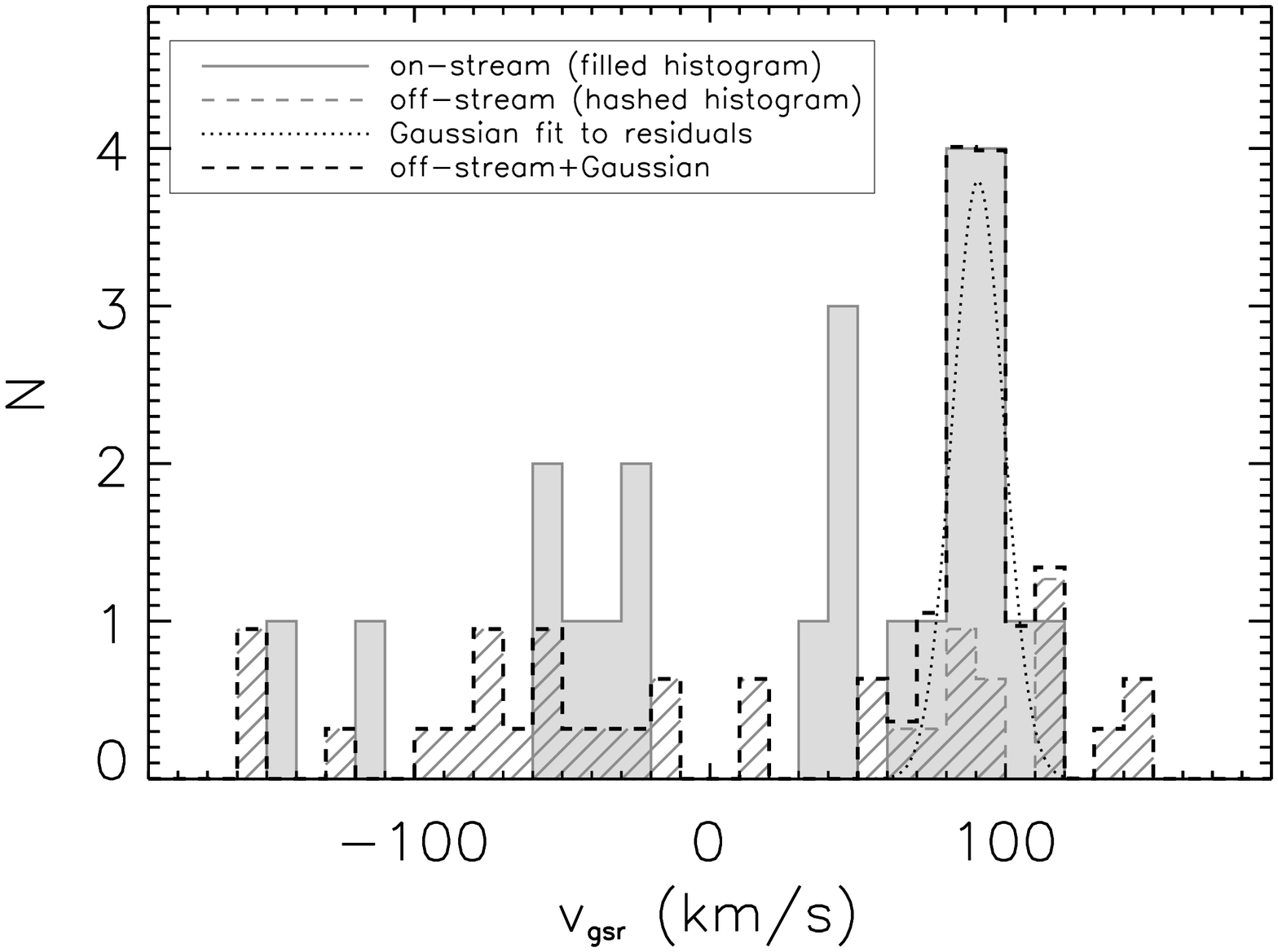}
\caption{{\it Left panel:} Histogram showing the samples that were depicted as colored (blue and red) points in Figure~\ref{spectra_CMD}, with an additional cut requiring stars to have $(g-r)_0 > 0.4$. Grey filled bars represent the on-stream sample and grey hashed bars represent stars in the off-stream field. The off-stream bins have been normalized so that the total number of stars with [Fe/H]$ > -1.5$ is equal in the on- and off-stream samples.
The on-stream stars consist of two different populations: the metal-rich peak centered on [Fe/H]~$\sim -0.7$ roughly matches the ``background'' stars of the off-stream field, while a more metal-poor population is evident between $-2.8 \lesssim$~[Fe/H]~$\lesssim -1.8$ only in the on-stream data. We fit a Gaussian to the residuals from subtracting the off-stream (``background'') from the on-stream histogram; the result is shown as a dotted curve centered at [Fe/H]=-2.24, with $\sigma_{[Fe/H]} = 0.32$. The sum of the off-stream counts and the Gaussian fit (dashed black line) matches the on-stream data well. {\it Right panel:} GSR-frame velocities of metal-poor ([Fe/H]$ < -1.5$) red giant stars in the on- and off-stream fields. Line styles, fill styles and scaling are as in the left panel. As in the metallicity plot, we subtracted the off-stream from the on-stream and fit a Gaussian to the residuals, this yields $\langle v_{\rm gsr} \rangle = 90.7~{\rm km~s^{-1}}$, with $\sigma_{\rm v} = 8.7~{\rm km~s^{-1}}$.
The apparent peak located at $v_{\rm gsr}\sim~50~{\rm km~s^{-1}}$ of stars appears to be a different population from the PSS. Three of the stars are separated by more than a magnitude from the PSS locus on the CMD and the final star has a velocity more than 6$\sigma$ from the measured value of the stream.
}
\label{hist}
\end{figure}

The surface gravity cut selects asymptotic giant branch (AGB) and blue horizontal branch (BHB) stars in addition to the RGB stars we are interested in finding. To 
further isolate RGB stars, we 
chose only stars with color $(g-r)_0 > 0.4$ from the velocity-selected sample shown in Figure~\ref{spectra_CMD} (light blue points), and histogrammed their metallicities in the upper panel of Figure~\ref{hist}.
Because the off-stream field has many more stars than the on-stream selection, we normalized the off-stream counts by the relative number of stars with [Fe/H]$ > -1.5$ (43 in the off-stream vs. 11 in the on-stream region).
The distributions of velocity-selected RGB candidate stars in the on- and 
off-stream fields are clearly different. The on-stream field has a metal-rich component at $-1.0 <$~[Fe/H]~$< -0.4$ that is consistent with the off-stream (i.e., background) stars, but also a different, 
more metal poor, population at [Fe/H]~$< -1.8$. Also shown in the left panel of Figure~\ref{hist} is a Gaussian that we fit to the residuals from subtracting the scaled off-stream counts from the on-stream data; this yields $\langle {\rm [Fe/H]} \rangle = -2.24$, with $\sigma_{\rm [Fe/H]} = 0.32$. Because this low-metallicity peak is not present in 
the off-stream fields, we identify this as the PSS metallicity signature; this is the reason for the 
metallicity selection of $-2.8 <$~[Fe/H]~$< -1.8$ for the red points in Figure~\ref{spectra_CMD}. 

We performed a similar procedure to find the velocity of stream candidates. The right panel of Figure~\ref{hist} shows velocities of metal-poor ([Fe/H]$ < -1.5$) red giants, with the off-stream number counts scaled down by a factor of 3.16 so that the number of red giants with [Fe/H]$ > -1.5$ is equal in the two samples. Again, we subtracted the scaled off-stream histogram from the on-stream data, and fit a Gaussian to the residuals. The fit (shown as a dotted curve in Figure~\ref{hist}) gives $\langle v_{\rm gsr} \rangle = 90.7~{\rm km~s^{-1}}$, with $\sigma_{\rm v} = 8.7~{\rm km~s^{-1}}$.

Having confirmed the presence of a population of stars having common velocity and metallicity, we 
next examine the photometric data in this region of the sky to look for additional evidence of the 
substructure. We selected
all stars with photometric data from the SDSS DR8 database,
within the on- and off-stream fields described above. We created CMD Hess diagrams 
for both the on- and off-stream fields using extinction-corrected $g_0$ magnitudes and $(g-r)_0$
colors. The Hess diagram for the off-stream field was normalized by dividing number counts in 
each bin by 8 to account for the $8\times$ larger area relative to the on-stream 
region. We subtracted the normalized off-stream Hess diagram from that of the on-stream field to 
create a difference plot. The residuals from this subtraction can be seen in the bottom left panel of 
Figure~\ref{spectra_CMD}. 

Figure~\ref{spectra_CMD} (lower panels) reveals an overdensity of stream stars between $0.2 < (g-r)_0 < 0.4$ and $20.5 < g_0 < 21.6$,
consistent with a population of F-turnoff stars from the stream whose velocity signature we have found in RGB stars. 
The turnoff magnitude is clearly fainter than that of the 28 kpc isochrone, and appears more likely to be between the 33-kpc and 38-kpc isochrones. Since the stars are in a narrow line across the sky, we surmise that the turnoff stars and spectroscopically selected RGB stars are part of an old, metal-poor tidal stream at a distance of $\sim35\pm3$ kpc.

\begin{deluxetable}{lrcrrrrrr}
\tabletypesize{\small}
\tablecaption{Pisces Stellar Stream Candidates \label{cpsk}}
\tablehead{\colhead{ra (deg)} & \colhead{dec (deg)} & \colhead{$V_{gsr}$ (km s$^{-1}$)} & \colhead{[Fe/H]} & \colhead{$\log {g}$} & \colhead{$g_0$} & \colhead{$(g-r)_0$} & \colhead{$(g-i)_0$} & \colhead{$(u-g)_0$}} 

\startdata
\\
24.18291 & 22.93643 & 113.5 & -2.54 & 2.30 & 18.14 & 0.59 & 0.84 & 1.23\\
24.21572 & 22.95978 & 93.1 & -1.87 & 2.92 & 18.98 & 0.53 & 0.73 & 1.15\\
23.80551 & 23.47830 & 96.7 & -2.37 & 2.72 & 18.89 & 0.50 & 0.67 & 0.98\\
23.92007 & 23.39030 & 85.5 & -2.13 & 2.18 & 18.71 & 0.52 & 0.73 & 1.22\\
23.24328 & 23.19339 & 80.9 & -2.12 & 3.70 & 17.84 & -0.14 & -0.26 & 1.28\\
23.57924 & 24.39110 & 101.7 & -2.65 & 1.69 & 19.12 & 0.56 & 0.76 & 1.21\\
23.41445 & 24.34511 & 93.1 & -2.07 & 3.09 & 19.10 & 0.52 & 0.72 & 1.12\\
23.78164 & 23.88001 & 77.4 & -2.38 & 1.85 & 18.74 & 0.55 & 0.76 & 1.23\\
24.15154 & 23.36385 & 88.0 & -2.35 & 1.09 & 17.91 & 0.62 & 0.89 & 1.23\\
23.91934 & 24.11852 & 82.5 & -1.95 & 3.12 & 18.70 & 0.53 & 0.78 & 1.13\\
23.82848 & 22.80306 & 94.4 & -1.91 & 2.08 & 17.34 & 0.65 & 0.91 & 1.44\\
\\

\enddata
\end{deluxetable}

\section{Discussion and Conclusion}

We have identified the kinematical and chemical signature of a likely stellar tidal stream based on 
spectra in an SDSS plate located at $l \approx 135^\circ$ and between $-39^\circ < b < -36^\circ$.  
Stars in this structure, which we dub the Pisces Stellar Stream (PSS), have line-of-sight velocities in 
the Galactic standard of rest frame between $77$~km~s$^{-1} < v_{\rm gsr} <$~114~km~s$^{-1}$.
We estimate the mean velocity for the 11 stars using a maximum likelihood method (e.g., Pryor \& Meylan 1993; Kleyna et al. 2002), and find $\langle v_{\rm gsr} \rangle = 95.6\pm3.1$ km~s$^{-1}$ and an intrinsic velocity
dispersion (accounting for individual velocity errors; the typical error for an individual star is $\sim7.0$~km~s$^{-1}$) of $\sigma_{\rm v,0} = 7.6\pm3.0$ km~s$^{-1}$. 
The metallicity of these 11 stars, determined from a Gaussian fit to the residuals after subtracting an off-stream field, is $\langle $[Fe/H]$\rangle = -2.24$ with a dispersion of $\sigma_
{\rm [Fe/H]} = 0.32$, which is only slightly larger than the SDSS metallicity measurement error 
of $\sim0.2$ dex \citep{2008AJ....136.2022L}.
We isolate the PSS in both a background-subtracted color-magnitude Hess 
diagram and a matched-filter star count map. From these, we derive a distance to the stream of 
$35\pm3$~kpc. No distance gradient was apparent over the short stretch of the stream we isolated. 
The mean position of the 11 spectroscopically confirmed stars is $(\alpha, \delta) = (23.82^\circ, 
23.53^\circ)$.

The Pisces stream is a small segment of the
structure identified as ``stream a'' by Grillmair~(2012) in a matched-filter star count map of the south Galactic cap. This structure has also been reported by Bonaca et 
al.~(2012) based on star count maps using a similar technique. Bonaca et al. found a distance of 
$26\pm4$ kpc to the stream, which they dubbed the ``Triangulum stream''. However, this distance 
was based on a metallicity of [Fe/H] = -1.0 estimated from isochrones; our measured spectroscopic 
metallicity of [Fe/H] = -2.2 requires a more distant, $\sim35$ kpc isochrone to match the stream in the 
CMD. Our finding that the stream is kinematically cold, narrow in width, and essentially mono-metallic, confirms the suggestions of both Grillmair~(2012) and Bonaca et al.~(2012) that this stream is the remnant of a globular cluster, or possibly an ultrafaint dwarf galaxy.

This stream is distinct in distance and line-of-sight velocity from other known substructures in this region of the sky. The Triangulum-Andromeda 
(TriAnd) stellar structure (Rocha-Pinto et al.~2004) is a large, diffuse substructure. TriAnd is $\sim$16-25~kpc from the Sun (Majewski~et~al.~2004), much closer than our 35~kpc estimate for the PSS.
Moreover, our mean 
velocity of $\langle v_{\rm gsr} \rangle = 95.6$ km~s$^{-1}$ at $l \approx 135^\circ$ is much higher 
than the $\sim30$ km s$^{-1}$ found by Rocha-Pinto et al. (2004) for TriAnd at this position. We thus consider 
an association between the Pisces stream and TriAnd to be unlikely. The PSS is also unlikely to be associated with the CPS, as the CPS has $v_{\rm gsr} \sim -60$ km s$^{-1}$ at $b \approx -38^\circ$, compared to our $\sim 90$ km s$^{-1}$ velocity for the Pisces stream at this position.

\acknowledgements

We thank the anonymous referee for insightful comments. This work was supported by NSF grants AST 09-37523, AST 10-09670, and AST 12-39904. Funding for SDSS-III has been provided by the Alfred P. Sloan Foundation, the Participating Institutions, the National Science Foundation, and the U.S. Department of Energy Office of Science. The SDSS-III web site is \url{http://www.sdss3.org/}.

\bibliographystyle{apj}

\begin{thebibliography}
\expandafter\ifx\csname natexlab\endcsname\relax\def\natexlab#1{#1}\fi

\bibitem[Belokurov et al.(2006)]{2006ApJ...642L.137B} Belokurov, V., 
Zucker, D.~B., Evans, N.~W., et al.\ 2006, \apjl, 642, L137 

\bibitem[Belokurov et al.(2007)]{2007ApJ...654..897B} Belokurov, V.,
Zucker, D.~B., Evans, N.~W., et al.\ 2007, \apj, 654, 897 

\bibitem[Bonaca et al.(2012)]{2012ApJ...760L...6B} Bonaca, A., Geha, M., 
\& Kallivayalil, N.\ 2012, \apjl, 760, L6 

\bibitem[Carlin et al.(2012)]{2012ApJ...753..145C} Carlin, J.~L., Yam, W., 
Casetti-Dinescu, D.~I., et al.\ 2012, \apj, 753, 145 

\bibitem[Duffau et al.(2006)]{2006ApJ...636L..97D} Duffau, S., Zinn, R., 
Vivas, A.~K., et al.\ 2006, \apjl, 636, L97 

\bibitem[Girardi et al.(2010)]{2010ApJ...724.1030G} Girardi, L., Williams, 
B.~F., Gilbert, K.~M., et al.\ 2010, \apj, 724, 1030 

\bibitem[Grillmair 
\& Johnson(2006)]{2006ApJ...639L..17G} Grillmair, C.~J., \& Johnson, R.\ 2006, \apjl, 639, L17 

\bibitem[Grillmair 
\& Dionatos(2006a)]{2006ApJ...641L..37G} Grillmair, C.~J., \& Dionatos, O.\ 2006a, \apjl, 641, L37 

\bibitem[Grillmair 
\& Dionatos(2006b)]{2006ApJ...643L..17G} Grillmair, C.~J., \& Dionatos, O.\ 2006b, \apjl, 643, L17 

\bibitem[Grillmair(2006)]{2006ApJ...651L..29G} Grillmair, C.~J.\ 2006, 
\apjl, 651, L29 

\bibitem[Grillmair(2009)]{2009ApJ...693.1118G} Grillmair, C.~J.\ 2009, 
\apj, 693, 1118 

\bibitem[Grillmair(2010)]{2010gama.conf..247G} Grillmair, C.~J.\ 2010, 
Galaxies and their Masks, 247 

\bibitem[Grillmair(2012)]{2012ASPC..458..219G} Grillmair, C.~J.\ 2012, 
Galactic Archaeology: Near-Field Cosmology and the Formation of the Milky 
Way, 458, 219 


\bibitem[Ibata et al.(2003)]{2003MNRAS.340L..21I} Ibata, R.~A., Irwin, 
M.~J., Lewis, G.~F., Ferguson, A.~M.~N., 
\& Tanvir, N.\ 2003, \mnras, 340, L21 


\bibitem[{{Kleyna} {et~al.}(2002){Kleyna}, {Wilkinson}, {Evans}, {Gilmore}, \&
  {Frayn}}]{kwe+02}
{Kleyna}, J., {Wilkinson}, M.~I., {Evans}, N.~W., {Gilmore}, G., \& {Frayn}, C.
  2002, \mnras, 330, 792

\bibitem[Koposov et al.(2010)]{2010ApJ...712..260K} Koposov, S.~E., Rix, 
H.-W., \& Hogg, D.~W.\ 2010, \apj, 712, 260 

\bibitem[Koposov et al.(2012)]{2012ApJ...750...80K} Koposov, S.~E., 
Belokurov, V., Evans, N.~W., et al.\ 2012, \apj, 750, 80 

\bibitem[Lee et al.(2008)]{2008AJ....136.2022L} Lee, Y.~S., Beers, T.~C., 
Sivarani, T., et al.\ 2008, \aj, 136, 2022 

\bibitem[Li et al.(2012)]{2012ApJ...757..151L} Li, J., Newberg, H.~J., 
Carlin, J.~L., et al.\ 2012, \apj, 757, 151 

\bibitem[Majewski et al.(2003)]{2003ApJ...599.1082M} Majewski, S.~R., 
Skrutskie, M.~F., Weinberg, M.~D., 
\& Ostheimer, J.~C.\ 2003, \apj, 599, 1082 

\bibitem[Majewski et al.(2004)]{2004ApJ...615..738M} Majewski, S.~R., 
Ostheimer, J.~C., Rocha-Pinto, H.~J., et al.\ 2004, \apj, 615, 738 

\bibitem[Majewski et al.(2012)]{2012ApJ...747L..37M} Majewski, S.~R., 
Nidever, D.~L., Smith, V.~V., et al.\ 2012, \apjl, 747, L37 

\bibitem[Marigo et 
al.(2008)]{2008A&A...482..883M} Marigo, P., Girardi, L., Bressan, A., et al.\ 2008, \aap, 482, 883 

\bibitem[Newberg et al.(2002)]{2002ApJ...569..245N} Newberg, H.~J., Yanny, 
B., Rockosi, C., et al.\ 2002, \apj, 569, 245 

\bibitem[Newberg et al.(2009)]{2009ApJ...700L..61N} Newberg, H.~J., Yanny, 
B., \& Willett, B.~A.\ 2009, \apjl, 700, L61 

\bibitem[{{Pryor} \& {Meylan}(1993)}]{pm93}
{Pryor}, C., \& {Meylan}, G. 1993, in Astronomical Society of the Pacific
  Conference Series, Vol.~50, Structure and Dynamics of Globular Clusters, ed.
  {S.~G.~Djorgovski \& G.~Meylan}, 357

\bibitem[Rocha-Pinto et al.(2004)]{2004ApJ...615..732R} Rocha-Pinto, H.~J., 
Majewski, S.~R., Skrutskie, M.~F., Crane, J.~D., 
\& Patterson, R.~J.\ 2004, \apj, 615, 732\

\bibitem[Rocha-Pinto(2010)]{2010IAUS..265..255R} Rocha-Pinto, H.~J.\ 2010, 
IAU Symposium, 265, 255 

\bibitem[Rockosi et al.(2002)]{2002AJ....124..349R} Rockosi, C.~M., 
Odenkirchen, M., Grebel, E.~K., et al.\ 2002, \aj, 124, 349 

\bibitem[Schlegel et al.(1998)]{1998ApJ...500..525S} Schlegel, D.~J., 
Finkbeiner, D.~P., \& Davis, M.\ 1998, \apj, 500, 525 

\bibitem[Sesar et al.(2012)]{2012ApJ...755..134S} Sesar, B., Cohen, J.~G., 
Levitan, D., et al.\ 2012, \apj, 755, 134 

\bibitem[Sharma et al.(2010)]{2010ApJ...722..750S} Sharma, S., Johnston, 
K.~V., Majewski, S.~R., et al.\ 2010, \apj, 722, 750 

\bibitem[Sheffield et al.(2012)]{2012ApJ...761..161S} Sheffield, A.~A., 
Majewski, S.~R., Johnston, K.~V., et al.\ 2012, \apj, 761, 161 


\bibitem[Vivas et al.(2001)]{2001ApJ...554L..33V} Vivas, A.~K., Zinn, R., 
Andrews, P., et al.\ 2001, \apjl, 554, L33 

\bibitem[Willett et al.(2009)]{2009ApJ...697..207W} Willett, B.~A., 
Newberg, H.~J., Zhang, H., Yanny, B., \& Beers, T.~C.\ 2009, \apj, 697, 207 

\bibitem[Yam et al.(2013)]{yam2013} Yam, W., Newberg, H.~J., Carlin, J.~L., Dumas, J., O'Malley, E., Martin, C., \& Newby, M. \ 2013, \apj, {\it submitted}

\bibitem[Yanny et al.(2003)]{2003ApJ...588..824Y} Yanny, B., Newberg, 
H.~J., Grebel, E.~K., et al.\ 2003, \apj, 588, 824 
  
\end{thebibliography}

\end{document}